\documentclass[useAMS,usenatbib,twocolappendix]{emulateapj}

\usepackage{ulem}
\usepackage{color}
\usepackage{graphics,graphicx}
\usepackage{times} 
\usepackage{amssymb}
\usepackage{amsmath}
\usepackage{natbib}
\usepackage{hyperref}
\usepackage{multirow}
\usepackage{ctable}
\usepackage{threeparttable}
\newif\ifAMStwofonts
\AMStwofontstrue

\def\xmm{{\it XMM-Newton}}

\def\epicpn{{EPIC-pn}}
\def\epicmos1{{EPIC-MOS1}}
\def\epicmos2{{EPIC-MOS2}}
\def\epicmos{{EPIC-MOS}}

\def\nustar{{\it NuSTAR}}

\def\m{{$\rm\thinspace m$}}

\def\pcmsq{\hbox{$\rm\thinspace cm^{-2}$}}
\def\H0{{\rm ~km~s^{-1}~Mpc^{-1}}}

\def\ergps{\hbox{erg~s$^{-1}$}}

\def\msun{\hbox{$M_{\odot}$}}

\def\chisq{{$\chi^{2}$}}

\def\xspecv{\hbox{\small XSPEC}\, v12.6.0f}

\def\nustardas{\rm {\small NUSTARDAS}}

\def\flx2xsp{\rm{\small FLX2XSP}}

\def\hendrics{\rm {\small HENDRICS}}
\def\pint{\rm {\small PINT}}

\def\grid25{\hbox{\rm{\small GRID25}}}

\def\simpl{\rm{\small SIMPL}}

\def\tbnew{\rm{\small TBNEW}}

\def\cutoffpl{\rm{\small CUTOFFPL}}
\def\fdcut{\rm{\small FDCUT}}
\def\npex{\rm{\small NPEX}}

\def\gabs{\rm{\small GABS}}

\def\eg{{\it e.g.}}

\def\ie{{\it i.e.~\/}}

\def\la{\mathrel{\hbox{\rlap{\hbox{\lower4pt\hbox{$\sim$}}}{\raise2pt\hbox{$<$}}}}}
\def\ga{\mathrel{\hbox{\rlap{\hbox{\lower4pt\hbox{$\sim$}}}{\raise2pt\hbox{$>$}}}}}

\def\d25{D$_{25}$}

\def\.25{0.25 keV\thinspace}

\def\rmag{$R_{\rm{M}}$}

\def\p13{NGC\,7793 P13}
\def\m82{M82 X-2}
\def\ulx1{NGC\,5907 ULX1}
\def\ngc{NGC\,300 ULX1}

\shorttitle{Cyclotron Scattering in NGC\,300 ULX}
\shortauthors{D.~J. Walton et al.}

\begin{document}

\title{A Potential Cyclotron Resonant Scattering
Feature in the ULX Pulsar NGC\,300 ULX1 seen by \textit{NuSTAR} and
\textit{XMM-Newton}}

\author{D. J. Walton\altaffilmark{1},
M. Bachetti\altaffilmark{2},
F. F\"urst\altaffilmark{3},
D. Barret\altaffilmark{4,5},
M. Brightman\altaffilmark{6},
A. C. Fabian\altaffilmark{1},
B. W. Grefenstette\altaffilmark{6},\\
F. A. Harrison\altaffilmark{6},
M. Heida\altaffilmark{6},
J. Kennea\altaffilmark{7},
P. Kosec\altaffilmark{1},
R. M. Lau\altaffilmark{6},
K. K. Madsen\altaffilmark{6},
M. J. Middleton\altaffilmark{8},
C. Pinto\altaffilmark{1},
J. F. Steiner\altaffilmark{9},\\
%D. Stern\altaffilmark{10},
N. Webb\altaffilmark{4,5} \\
}
\affil{ \\
$^{1}$ Institute of Astronomy, Cambridge University, Madingley Road, Cambridge, CB3 0HA \\ 
$^{2}$ INAF/Osservatorio Astronomico di Cagliari, via della Scienza 5, I-09047 Selargius (CA), Italy \\
$^{3}$ European Space Astronomy Centre (ESA/ESAC), Operations Department, Villanueva de la Ca\~nada (Madrid), Spain \\
$^{4}$ Universite de Toulouse; UPS-OMP; IRAP; Toulouse, France \\
$^{5}$ CNRS; IRAP; 9 Av. colonel Roche, BP 44346, F-31028 Toulouse cedex 4, France \\
$^{6}$ Cahill Center for Astrophysics, California Institute of Technology, 1216 E. California Blvd, Pasadena, CA 91125, USA \\
$^{7}$ Department of Astronomy and Astrophysics, Pennsylvania State University, 525 Davey Lab, University Park, PA 16802, USA \\
$^{8}$ Department of Physics and Astronomy, University of Southampton, Highfield, Southampton SO17 1BJ, UK \\
$^{9}$ MIT Kavli Institute for Astrophysics and Space Research, MIT, 70 Vassar Street, Cambridge, MA 02139, USA \\
%$^{10}$ Jet Propulsion Laboratory, California Institute of Technology, Pasadena, CA 91109, USA \\
}

\begin{abstract}
Based on phase-resolved broadband spectroscopy using \xmm\ and \nustar, we report
on a potential cyclotron resonant scattering feature at $E\sim13$\,keV in the pulsed
spectrum of the recently discoverd ULX pulsar \ngc. If this interpretation is correct, the
implied magnetic field of the central neutron star is $B\sim10^{12}$\,G (assuming
scattering off electrons), similar to that estimated from the observed spin-up of the star,
and also similar to known Galactic X-ray pulsars. We discuss the implications of this
result for the connection between \ngc\ and the other known ULX pulsars, particularly in
light of the recent discovery of a likely proton Cyclotron line in another ULX, M51 ULX-8.
\end{abstract}

\begin{keywords}
{Neutron Stars -- X-rays: binaries -- X-rays: individual (NGC\,300 ULX1)}
\vspace{-0.5cm}
\end{keywords}

\section{Introduction}

Ultraluminous X-ray sources (ULXs) are off-nuclear sources that appear to radiate in
excess of the Eddington limit for the standard $\sim$10\,\msun\ stellar remnant black
holes seen in Galactic X-ray binaries (\ie $L_{\rm{X}}>10^{39}$\,\ergps; see
\citealt{Kaaret17rev} for a recent review). The discovery that at least some ULXs are
powered by accreting pulsars brought about a paradigm shift in our understanding of
this exotic population. These neutron stars are the most extreme persistent accretors
known, with apparent luminosities up to $\sim$500 times the Eddington limit for a
standard neutron star mass of 1.4\,\msun\ ($\sim$2$\times$10$^{38}$\,\ergps). Until
recently only three such sources were known: \m82 (\citealt{Bachetti14nat}), \p13
(\citealt{Fuerst16p13, Israel17p13}) and \ulx1 (\citealt{Israel17}). In addition, although
pulsations have not been seen from this source, \cite{Brightman18} report the likely
detection of a cyclotron resonant scattering feature (CRSF) in the spectrum of a ULX
in M51, which would require this to be another neutron star accretor. However, the
similarity of these sources to the broader ULX population has led to speculation that
the neutron star accretors may dominate the demographics of ULXs (\citealt{Pintore17,
Koliopanos17, Walton18p13, Walton18ulxBB}). The recent discovery of a fourth ULX
pulsar in NGC\,300 by \cite{Carpano18} provides further evidence that neutron stars
may be common among ULXs.

\begin{figure*}
\begin{center}
\hspace*{-0.3cm}
\rotatebox{0}{
{\includegraphics[width=230pt]{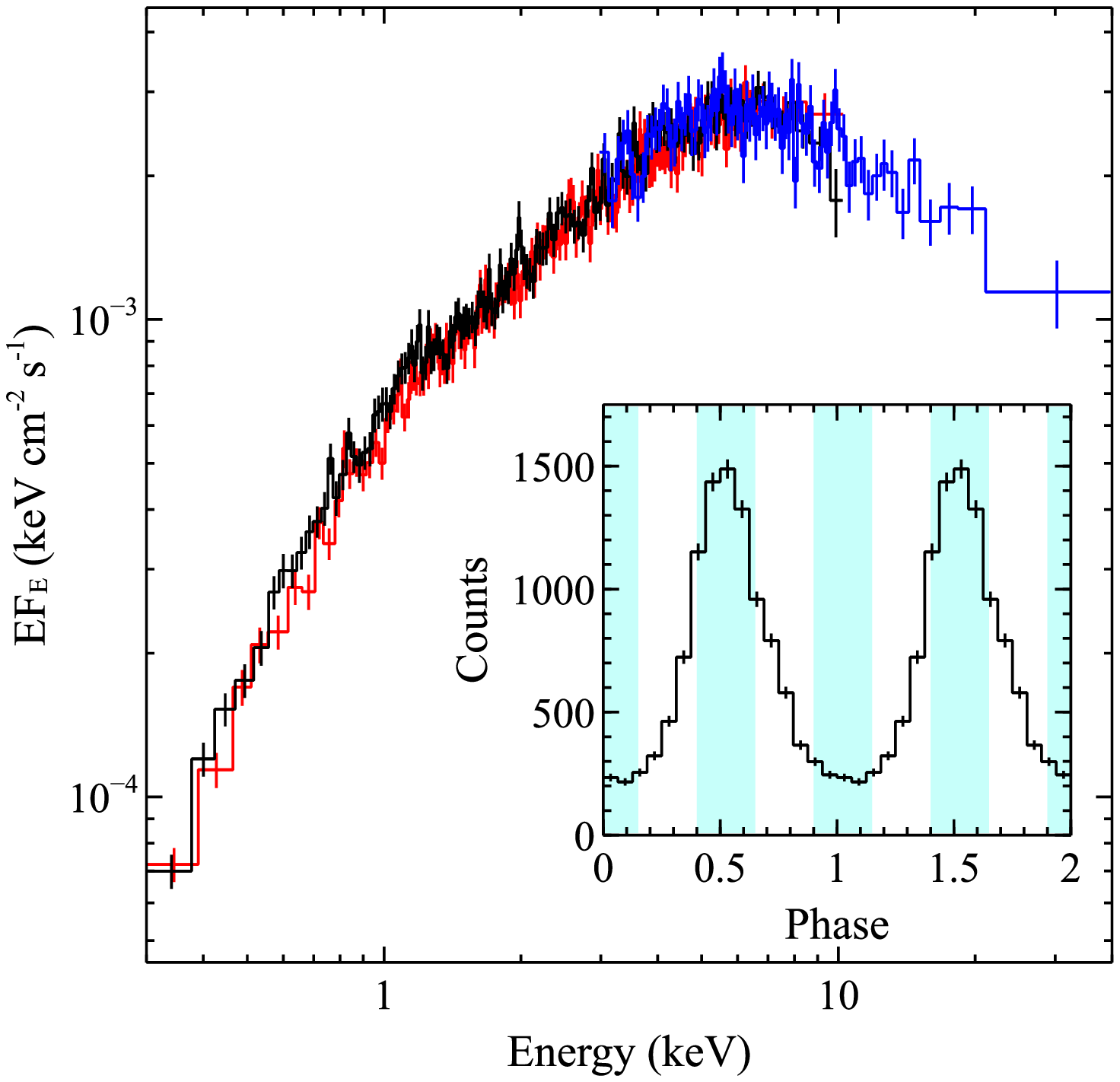}}
}
\hspace*{1cm}
\rotatebox{0}{
{\includegraphics[width=230pt]{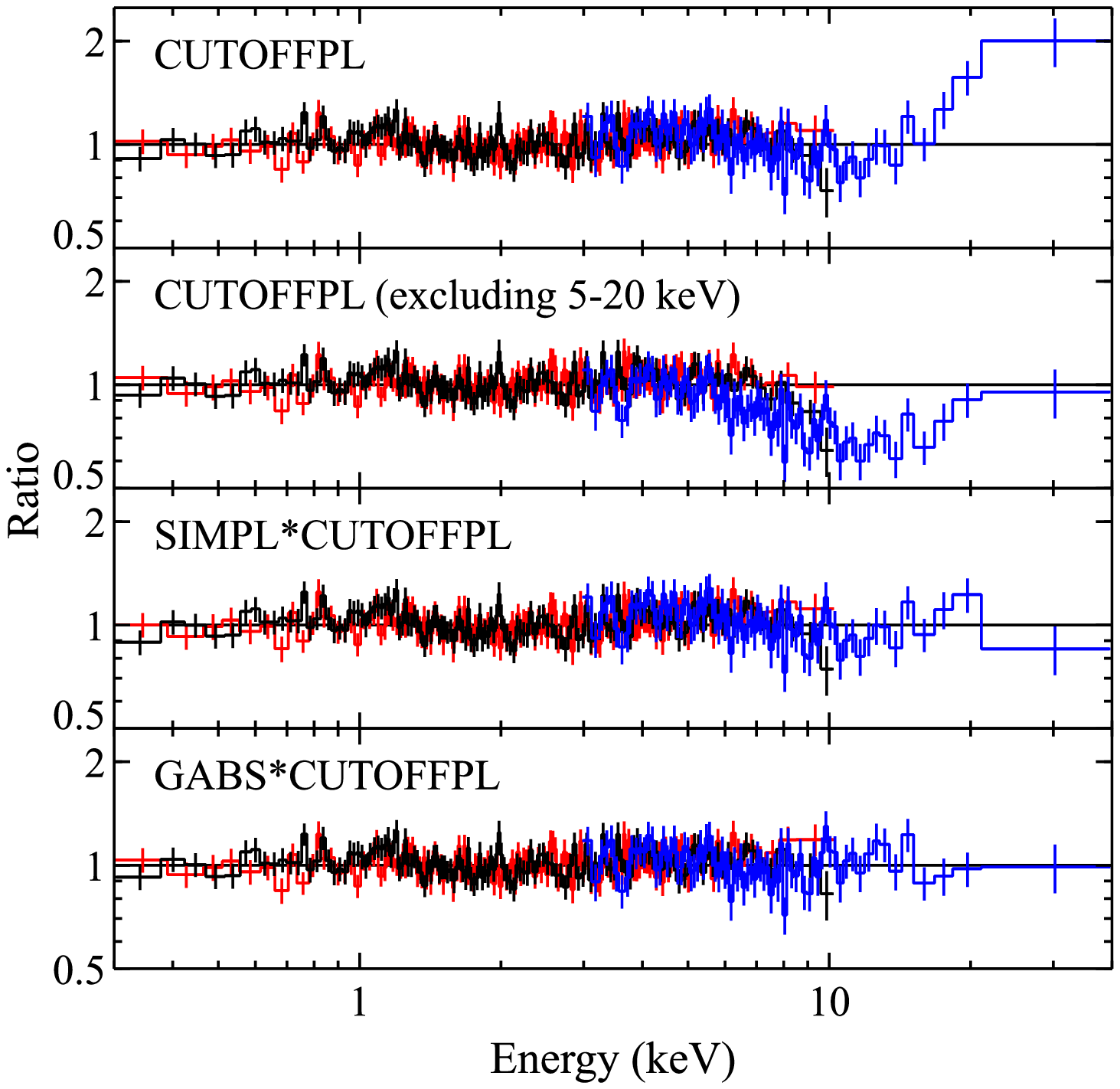}}
}
\end{center}
\caption{
\textit{Left:} the pulsed spectrum of \ngc\ observed in 2016 by \xmm\ (\epicpn\ in black,
\epicmos\ in red) and \nustar\ (blue; while we fit FPMA and FPMB separately in our
analysis, for plotting purposes we combine the two modules), unfolded through a
model that is constant with energy. \textit{Inset:} the pulse profile in the 3--40\,keV
\nustar\ band; the shaded regions indicate the periods from which data were
extracted to produce the pulsed spectrum in the main panel. \textit{Right:} data/model
ratios for some of the models considered (see Section \ref{sec_pulse}). The data have
been rebinned for visual clarity.}
\label{fig_spec}
\vspace{0.4cm} 
\end{figure*}

\ngc\ was originally identified as a supernova candidate in 2010 after a strong optical
outburst, and given the identifier SN 2010da (\citealt{Monard10ngc300}). However,
the discovery of a moderately bright ($L_{\rm{X}}\sim10^{38}$\,\ergps), recurring
X-ray counterpart and its subsequent rebrightening in the infrared $\sim$2000 days
after the optical outburst ultimately helped to confirm \ngc\ to be a high-mass X-ray
binary (\citealt{Binder16, Lau16ngc300, Villar16}). Observations in December 2016
revealed a further X-ray outburst from this source, in which it reached observed ULX
luminosities ($L_{\rm{X}}\sim3\times10^{39}$\,\ergps; \citealt{Carpano18}), and the
detection of a $\sim$32\,s pulse period from this epoch confirmed the accretor as a
neutron star (and the fourth ULX pulsar). Subsequent X-ray monitoring revealed the
neutron star to be undergoing an extreme rate of spin-up, with $\dot{P} > 10^{-7}$\,s/s,
resulting in a pulse period of $\sim$20\,s only $\sim$15 months after the 2016
observations in which the pulsations were initially discovered (\citealt{Kennea18atel300,
Bachetti18atel300}). Similar to the other ULX pulsars the pulse profile appears to be
broad and relatively sinusoidal (\citealt{Carpano18}), but one of the remarkable aspects
of this particular source is that its pulsed fraction is extremely high, reaching $>$75\% at
energies $>$2\,keV. Finally, there is also now evidence that this source is launching an
extreme and variable X-ray wind (\citealt{Kosec18}), consistent with the basic
expectation for super-Eddington accretion (\citealt{Shakura73, Poutanen07})
and similar to other ULXs (\citealt{Pinto16nat, Walton16ufo, Kosec18ulx}).

Here we present a broadband X-ray spectral analysis of \ngc, in which we find
potential evidence for a CRSF in the 2016 \xmm+\nustar\ data.

\section{Observations and Data Reduction}
\label{sec_red}

\nustar\ (\citealt{NUSTAR}) and \xmm\ (\citealt{XMM}) performed a coordinated
observation of the galaxy NGC\,300 starting on 17 December 2016 (\xmm\ OBSIDs
0791010101 and 0791010101, taken over back-to-back orbits, and \nustar\ OBSID
30202035002). Although the primary target of the observation was the Wolf-Rayet
X-ray binary NGC\,300 X-1, the observation serendipitously caught an outburst of \ngc.
The two sources are separated by $\sim$70$''$, and during these observations ULX1
is more than an order of magnitude brighter than X-1. Our reduction of the \nustar\
data follows the procedure outlined in \cite{Kosec18}, and we extract both the standard
`science' and the `spacecraft science' data (\citealt{Walton16cyg}). For \xmm\ we just
focus on the data from the \epicpn\ and \epicmos\ detectors (\citealt{XMM_PN,
XMM_MOS}). Our reduction again largely follows \cite{Kosec18}, although we use a
slightly smaller region of 30$''$ radius to extract source products to reduce the
background. The total good exposures are 140\,ks and 190\,ks for the \epicpn\ and
\epicmos\ detectors, and 180\,ks for the \nustar\ FPMA/B modules, respectively. In all
cases, the cleaned event files were barycentred using the DE-200 ephemeris.

\vspace{0.1cm}
\section{The Pulsed Spectrum of NGC\,300 ULX1}
\label{sec_pulse}

Our analysis focuses on the broadband X-ray spectrum of the pulsed emission from the
accretion column in \ngc. Model fits are performed with \xspecv\ (\citealt{xspec}), and
unless stated otherwise uncertainties on the spectral parameters are quoted at the 90\%
confidence level for one interesting parameter. All models presented include a Galactic
absorption component with a fixed column of $N_{\rm{H,Gal}}=4.2\times10^{20}$\,\pcmsq\
(\citealt{NH}), and we also allow for absorption intrinsic to the source at the redshift of
NGC\,300 ($N_{\rm{H;int}}$; $z=0.00048$). Both absorption components are modelled
with the \tbnew\ absorption code, and we use the cross-sections of \cite{Verner96} and
the abundance set presented in \cite{tbabs}. We also allow for cross-calibration
uncertainties between the different detectors by including multiplicative constants that
are allowed to float between the datasets, fixing FPMA at unity. These are always within
$\sim$12\% of unity, as expected (\citealt{NUSTARcal}).

We isolate the spectrum of the pulsed component following the approach taken in our
recent analyses of \m82, \p13\ and \ulx1\ (\citealt{Brightman16m82a, Walton18p13,
Walton18ulxBB}). In brief, we phase-tag the cleaned event files, and extract spectra
from the brightest and the faintest quarters of the pulse cycle (the pulse-profile of \ngc\
is broad and nearly sinusoidal, fairly similar to the other known ULX pulsars; see Figure
\ref{fig_spec}). We then subtract the latter from the former (\ie ``pulse-on"$-$``pulse-off").
The timing solution used to phase-tag the events combines a pulse frequency of
$\nu=0.0315275(3)$\,Hz (at MJD 57738.65732) and a strong frequency derivative of
$\dot{\nu}=5.535(2)\times10^{-10}$\,Hz~s$^{-1}$ (the parentheses indicate the
1$\sigma$ error on the last digit). This was derived using a combination of the \hendrics\
(version 0.4rc1; \citealt{MaLTPyNT}) and \pint\footnote{https://github.com/nanograv/pint}
software packages, and is consistent with \cite{Carpano18}. The spectra extracted are
rebinned to have a minimum of 25 counts per energy bin to allow the use of \chisq\
statistics, and we fit the data over the $\sim$0.3--40\,keV energy range.

The pulsed spectrum is shown in Figure \ref{fig_spec} (left panel). We initially fit these
data with a \cutoffpl\ model, which provides an excellent description of the pulsed
emission in the other ULX pulsars. However, we find that this simple model can not
successfully fit the data for \ngc. Although the fit is not terrible in a statistical sense,
with $\chi^{2}=1669$ for 1597 degrees of freedom (DoF), the model leaves a clear
excess of emission at the highest energies; the data/model ratios for a number of the
models considered here are also shown in Figure \ref{fig_spec} (right panels). This is
reminiscent of the hard excesses ubiquitously seen in the average spectra of the
broadband ULX sample (\citealt{Walton18ulxBB}), including the ULX pulsars \p13\ and
\ulx1. However, a critical difference is that in the ULX pulsars (and potentially the rest
of the ULX population as well) these excesses in the average spectra are related to
the presence of the pulsed emission from the accretion column (\citealt{Walton18p13,
Walton18ulxBB}). Here, we have already isolated this pulsed emission.

Nevertheless, it is possible that the pulsed continuum in this particular case is more
complex than a simple \cutoffpl\ model can describe. We first attempt to fit these data
with other phenomenological models typically used for the high-energy emission in
neutron star X-ray binaries, which allow for more complex spectral shapes: \fdcut\ (a
cutoff powerlaw model which includes both the folding energy, $E_{\rm{fold}}$, and the
energy at which this folding begins to act, $E_{\rm{cut}}$, as free parameters;
\citealt{fdcut}) and \npex\ (a combination of two \cutoffpl\ models with $E_{\rm{fold}}$
linked between the two; traditionally the slope of one component is free to vary, while
the other is fixed to $\Gamma=-2$; \citealt{npex}). However, we find that both of
these models result in similar hard excesses as the \cutoffpl\ model (and the fits are
statistically similar, with $\chi^{2}/{\rm{DoF}}=1672/1596$ and 1663/1596 for \fdcut\
and \npex, respectively). The curvature in the $\sim$5--10\,keV band is too strong for
any smoothly varying single-component model to account for the highest energies
probed by \nustar, and a second model component is clearly required to fit the data.

\begin{table}
  \caption{Best fit parameters obtained for the pulsed spectrum with the models
  including a high-energy powerlaw tail}
\vspace{-0.3cm}
\begin{center}
%\hspace*{-0.15cm}
\begin{tabular}{c c c c c}
\hline
\hline
\\[-0.1cm]
Parameter & \multicolumn{2}{c}{Model Combination} \\
\\[-0.2cm]
& \simpl$\otimes$\cutoffpl & \simpl$\otimes$\fdcut\ \\
\\[-0.2cm]
\hline
\hline
\\[-0.1cm]
$N_{\rm{H; int}}$ ($10^{20}$\,cm$^{-2}$) & $2.3^{+1.0}_{-1.1}$ & $7.6^{+1.1}_{-1.3}$ \\
\\[-0.2cm]
$\Gamma$ & $0.72^{+0.06}_{-0.05}$ & $1.20^{+0.04}_{-0.06}$ \\
\\[-0.2cm]
$E_{\rm{cut}}$ (keV) & -- & $5.9^{+1.1}_{-0.4}$ \\
\\[-0.2cm]
$E_{\rm{fold}}$ (keV) & $4.9^{+0.5}_{-0.4}$ & $0.8 \pm 0.5$ \\
\\[-0.2cm]
Norm ($10^{-3}$) & $1.7^{+0.3}_{-0.6}$ & $0.8^{+0.7}_{-0.2}$ \\
\\[-0.2cm]
$\Gamma_{\rm{SIMPL}}$ & $<2.8$ & $2.7^{+0.1}_{-0.3}$ \\
\\[-0.2cm]
$f_{\rm{sc}}$ (\%) & $43^{+9}_{-39}$ & $>52$ \\
\\[-0.2cm]
\hline
\\[-0.15cm]
\chisq/DoF & 1654/1595 & 1588/1594 \\
\\[-0.2cm]
\hline
\hline
\end{tabular}
\label{tab_param_simpl}
\end{center}
\end{table}

Similar to our early work on the average ULX spectra, we first attempt to account for
this high-energy excess by allowing for a further high-energy powerlaw tail using
an additional \simpl\ component (\citealt{SIMPL}; note that this model has a lower limit
on the photon index of $\Gamma_{\rm{SIMPL}}\geq1$). In these subsequent fits, we
find that for the \npex\ model the normalisation of the $\Gamma=-2$ runs to zero,
making the model indistinguishable from the simpler \cutoffpl\ case, so from this point
we only report the results from the latter. The quality of fit depends on the choice of
continuum model (the results are presented in Table \ref{tab_param_simpl}), but in
both cases considered the addition of the \simpl\ component provides a significant
improvement ($\Delta\chi^{2}\geq15$ for two additional free parameters, giving a
chance improvement probability of 0.004 according to the Akaike Information Criterion),
and the high-energy data are much better described (see Figure \ref{fig_spec}).
However, when treating the hard excess as a powerlaw tail, the sharper curvature that
\fdcut\ can provide in the $\sim$5--10\,keV band is preferred.

\begin{table}
  \caption{Best fit parameters obtained for the pulsed spectrum with the models
  including a Gaussian CRSF}
\vspace{-0.3cm}
\begin{center}
%\hspace*{-0.15cm}
\begin{tabular}{c c c c c}
\hline
\hline
\\[-0.2cm]
Parameter & \multicolumn{2}{c}{Model Combination} \\
\\[-0.2cm]
& \gabs$\times$\cutoffpl\ & \gabs$\times$\fdcut\ \\
\\[-0.2cm]
\hline
\hline
\\[-0.1cm]
$N_{\rm{H; int}}$ ($10^{20}$\,cm$^{-2}$) & $4.0 \pm 1.0$ & $6.3^{+1.1}_{-1.3}$ \\
\\[-0.2cm]
$\Gamma$ & $0.88 \pm 0.04$ & $1.11^{+0.05}_{-0.08}$ \\
\\[-0.2cm]
$E_{\rm{cut}}$ (keV) & -- & $11.2^{+5.5}_{-7.1}$ \\
\\[-0.2cm]
$E_{\rm{fold}}$ (keV) & $7.4^{+0.7}_{-0.6}$ & $4.9^{+1.2}_{-1.8}$ \\
\\[-0.2cm]
Norm ($10^{-3}$) & $0.96 \pm 0.04$ & $1.0^{+0.5}_{-1.2}$ \\
\\[-0.2cm]
$E_{\rm{CRSF}}$ (keV) & $12.8^{+1.0}_{-0.9}$ & $12.8^{+1.1}_{-0.9}$ \\
\\[-0.2cm]
$\sigma_{\rm{CRSF}}$ (keV) & $3.1^{+0.8}_{-0.7}$ & $3.9^{+1.1}_{-0.9}$ \\
\\[-0.2cm]
$d_{\rm{CRSF}}$ & $3.5^{+1.7}_{-1.2}$ & $7.6^{+8.8}_{-3.5}$ \\
\\[-0.2cm]
\hline
\\[-0.15cm]
\chisq/DoF & 1607/1594 & 1593/1593 \\
\\[-0.2cm]
\hline
\hline
\end{tabular}
\label{tab_param_CRSF}
\end{center}
\end{table}

\begin{figure}
\begin{center}
\hspace*{-0.4cm}
\rotatebox{0}{
{\includegraphics[width=230pt]{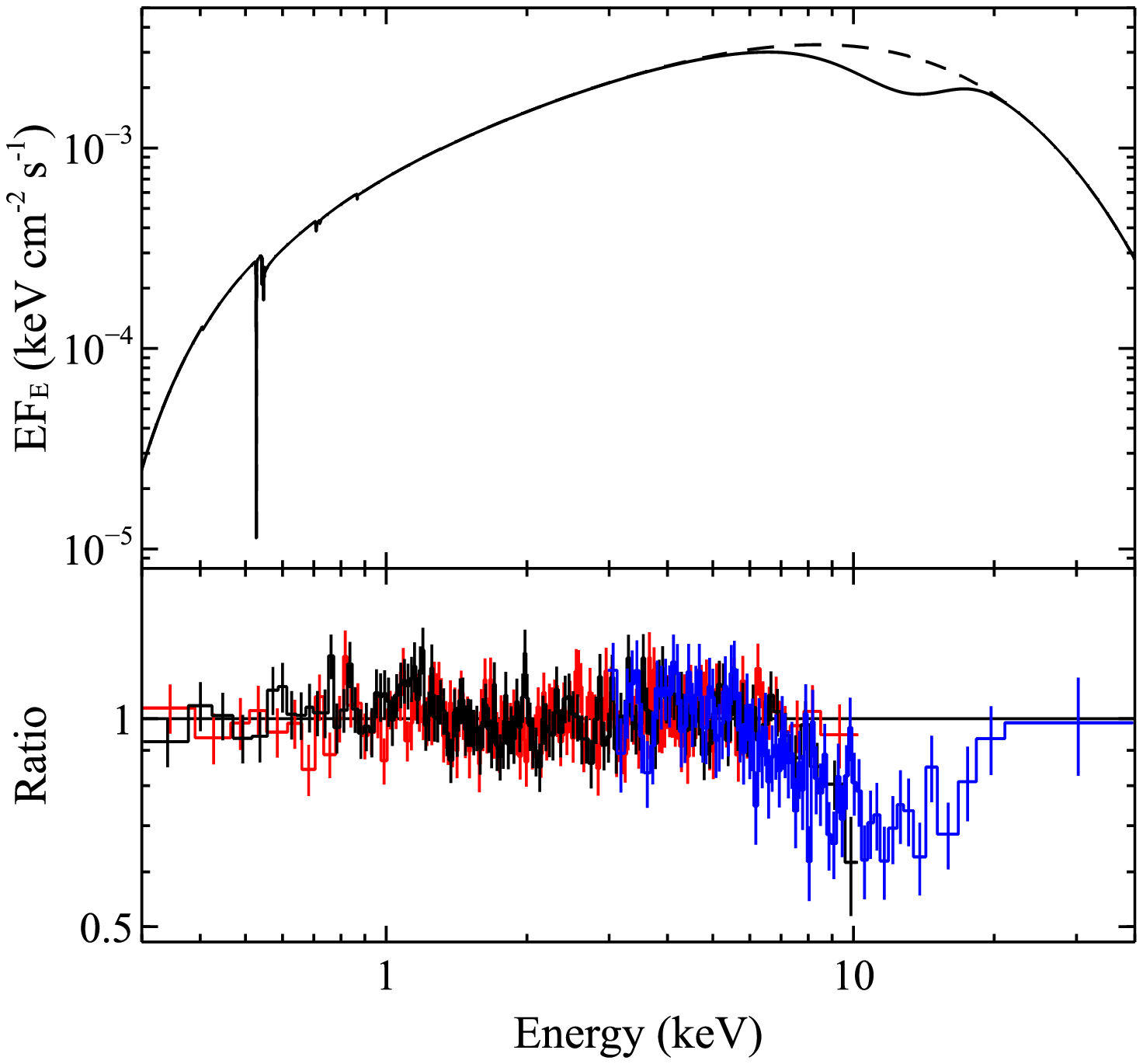}}
}
\end{center}
\caption{Top: the best-fit model combining GABS$\times$CUTOFFPL for the pulsed
spectrum of \ngc. The solid line shows the full model, and the dashed line shows the
intrinsic continuum after removal of the CRSF. Bottom: data/model ratio after the
removal of the CRSF, showing the best-fit line profile. This agrees well with that
inferred in Figure \ref{fig_spec} (the same data are shown).
}
\label{fig_cyclabs}
\vspace{0.2cm}
\end{figure}

\begin{figure*}
\begin{center}
\hspace*{-0.3cm}
\rotatebox{0}{
{\includegraphics[width=230pt]{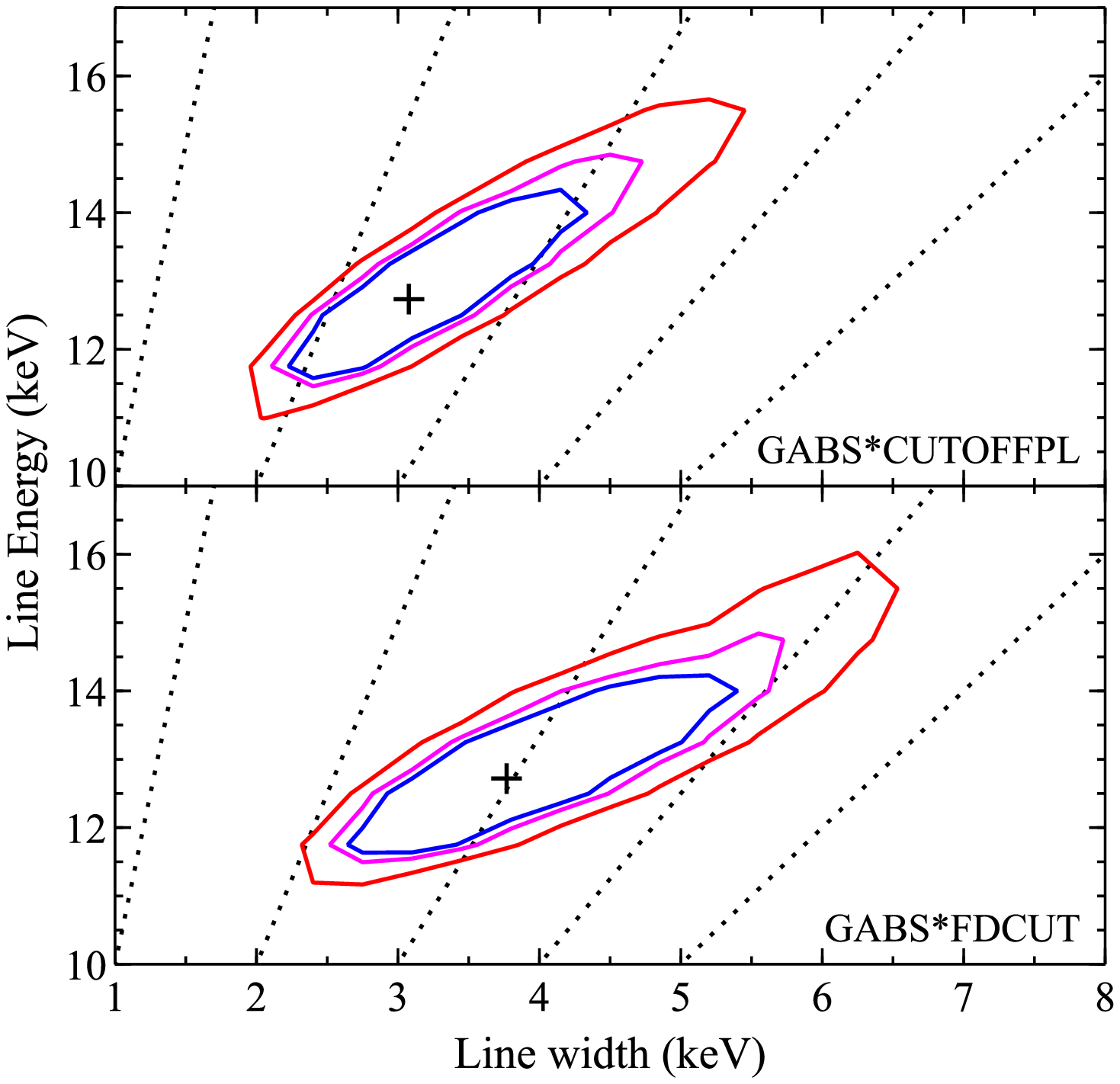}}
}
\hspace{1cm}
\rotatebox{0}{
{\includegraphics[width=230pt]{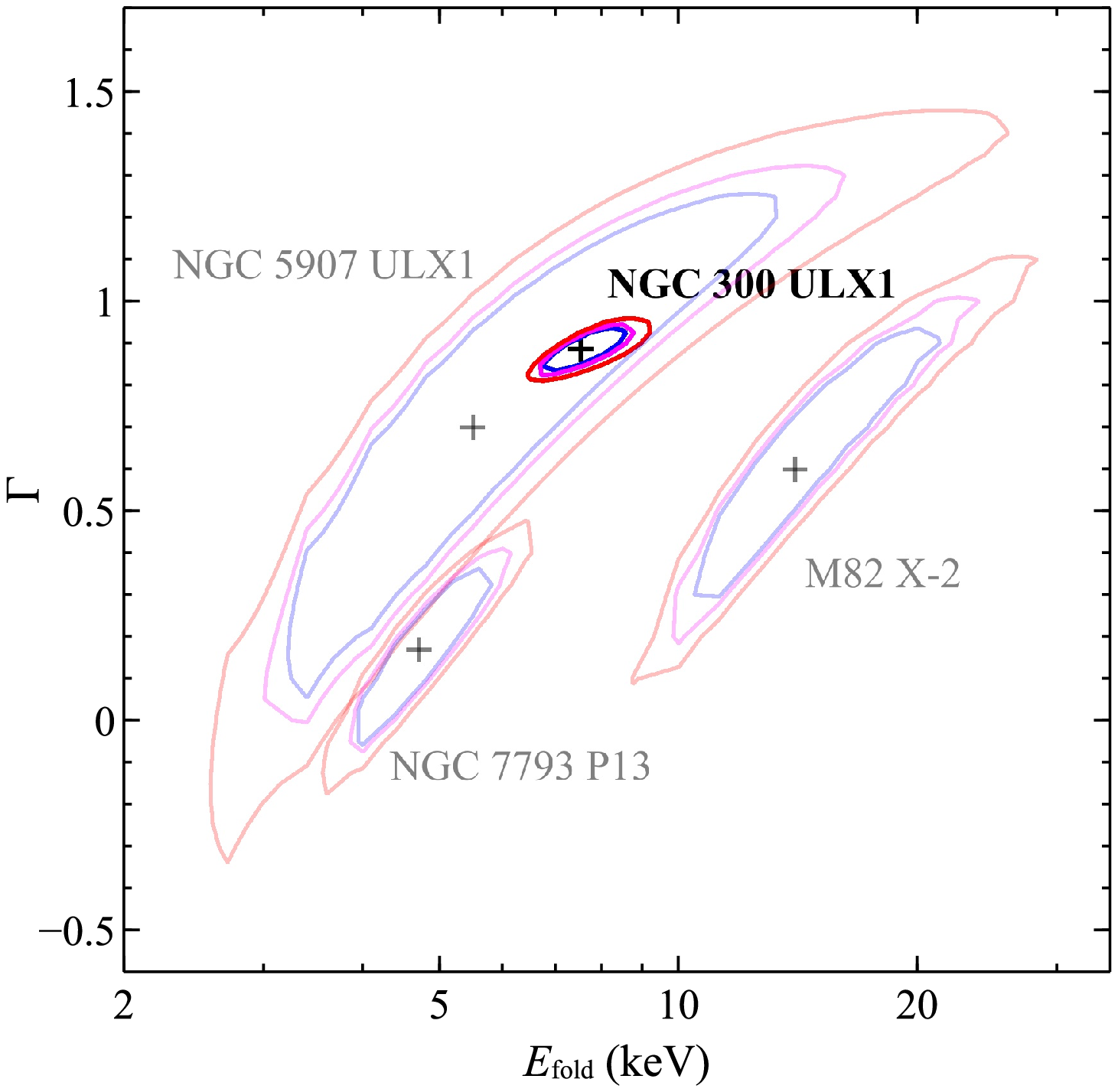}}
}
\end{center}
\caption{\textit{Left:} Confidence contours for the line energy and width for the
CUTOFFPL (top) and FDCUT (bottom) continuum models. The dotted lines show
$\sigma_{\rm{CRSF}}/E_{\rm{CRSF}}$ = 0.1, 0.2, 0.3, 0.4 and 0.5 (from left to right).
\textit{Right:} a comparison of the constraints on $\Gamma$ and $E_{\rm{fold}}$ for the
pulsed spectrum of \ngc\ with the GABS*CUTOFFPL (solid contours) model with the
constraints for the other known ULX pulsars (transparent contours). In each case, the
90, 95 and 99\% confidence contours for 2 parameters of interest are shown in blue,
magenta and red, respectively.
}
\label{fig_gamEc}
\vspace{0.3cm}
\end{figure*}

However, given that there is currently no evidence for similar hard excesses in the
pulsed emission from the other ULX pulsars, we also explore another possibility in
which the residuals seen in Figure \ref{fig_spec} are not caused by a real hard excess,
but are actually an indication of a broad cyclotron line causing a deficit in the
$\sim$5--20\,keV band (see Figure \ref{fig_spec}). We therefore replace \simpl\ with
\gabs, which provides a multiplicative line with a Gaussian optical depth profile (energy
$E$, width $\sigma$, depth $d$) and is often used to describe CRSF features (\eg\
\citealt{Staubert07, Fuerst14, Fuerst14vela, Jaisawal16}). For both of the continuum
models presented, this provides a similarly excellent description of the data, and a
significant improvement over the baseline continuum fits ($\Delta\chi^{2}\geq60$ for
three additional parameters; Table \ref{tab_param_CRSF}). We show the line profile
inferred with the \cutoffpl\ continuum in Figure \ref{fig_cyclabs}. The line parameters
are consistent for the two continuum models (see Figure \ref{fig_gamEc}), and there is
particularly good agreement over a line energy of $E_{\rm{CRSF}}=12.8$\,keV. In the
case of the \cutoffpl\ continuum, the CRSF provides a superior fit to the broadband data
than the model including \simpl, while for the \fdcut\ continuum the CRSF and \simpl\
fits are statistically very similar. We also show the contours of the baseline continuum
parameters for the \cutoffpl\ model for comparison with the other ULX pulsars in Figure
\ref{fig_gamEc}; aside from the additional complexity in the $5-20$\,keV band, the
pulsed spectrum of \ngc\ appears broadly similar to that seen from \m82, \p13\ and \ulx1.

\vspace{0.2cm}
\section{Discussion and Conclusions}
\label{sec_dis}

The pulsed spectrum from the 2016 \xmm+\nustar\ observation of the ULX pulsar \ngc\
-- the latest addition to the sample of ULX pulsars (\citealt{Carpano18}) -- cannot be fit
with simple, single-component continuum models. The curvature in the $\sim$5--15\,keV
band is quite strong, and there is more emission above $\sim$20\,keV than any smoothly
varying model that adequately fits this curvature can account for. In addition to a curved
baseline continuum, the data can be well explained with either an additional high-energy
powerlaw tail, or a cyclotron scattering feature at $E_{\rm{CRSF}}=12.8$\,keV imprinted
on top of an otherwise simple continuum. The other known ULX pulsars do not currently
show any evidence for a hard excess in their pulsed spectra that could also be related to
the presence of a high-energy powerlaw tail (\citealt{Brightman16m82a, Walton18p13,
Walton18ulxBB}). In addition, when invoking an additional powerlaw tail, $E_{\rm{fold}}$
is abnormally low when compared with Galactic X-ray pulsars for the \fdcut\ model
preferred in this scenario (\eg\ \citealt{Fuerst14vela, Vybornov17}). Therefore, although
we cannot rule out the possibility that the intrinsic pulsed continuum is more complex in
this case, we prefer the CRSF interpretation, and will focus our discussion on this
solution.

% Other potential fdcut Efold refs: Fuerst+13, Fuerst+15, Fuerst+18, Hemphill+15

The potential detection of a CRSF is significant, as these features offer the most robust
measure of the magnetic field ($B$) close to the surface of the neutron star, as the
energy of the fundamental electron CRSF (in keV) is given by
$E_{\rm{CRSF}}=11.57\times{B_{12}}(1+z_{\rm{grav}})$, assuming electron scattering.
Here, $B_{12}$ is the  magnetic field strength in units of $10^{12}$\,G, and
$z_{\rm{grav}}$ is the gravitational redshift of the line-forming region. Assuming the
potential CRSF is indeed the fundamental, this solution would therefore imply
$B\sim10^{12}$\,G (as $z_{\rm{grav}}\leq0.25$, since the line must be formed at or
beyond the neutron star surface). Critically, this is remarkably similar to the B-field
estimated from the spin-up of \ngc\ ($B\sim3\times10^{12}$\,G; \citealt{Carpano18}),
particularly given the uncertainties in both of these calculations (the exact position of
the line-forming region is not formally well known, \citealt{Poutanen13}, and since
$B\propto(\dot{P}/P)^{7/2}$ any errors on $P$ and $\dot{P}$ are magnified when
estimating $B$ with this method). This is also quite similar to the magnetic fields
inferred for typical Galactic X-ray pulsars (see \citealt{Caballero12rev} for a review).
Furthermore, the width of the line inferred in \ngc\ relative to its energy is similar to the
electron CRSF (eCRSF) features seen in other luminous X-ray pulsars (where such
features have been detected). Both V0332+53 and SMC X-2 show
$\sigma_{\rm{CRSF}}/E_{\rm{CRSF}}\sim0.2$ (\citealt{Tsygankov06, Jaisawal16}), and
for \ngc\ we find $0.2\lesssim\sigma_{\rm{CRSF}}/E_{\rm{CRSF}}\lesssim0.4$
(combining the constraints for the continuum models considered; see Figure
\ref{fig_gamEc}). As an aside, we note that since the line is broad, it should not
significantly impact the detection of the wind reported by \cite{Kosec18}.

The only other probable CRSF reported from a ULX to date comes from M51 ULX-8,
where \cite{Brightman18} report a narrow ($\sigma<0.2$\,keV) absorption feature
at 4.5\,keV. Although a firm identification is not possible, the preferred interpretation for
M51 ULX-8 is a proton CRSF (pCRSF), rather than an eCRSF. If correct, this would
imply an extreme (magnetar-like) magnetic field of $B\sim7\times10^{14}$\,G. This
identification was based on the ratio of the width and energy of the line, 
$\sigma_{\rm{CRSF}}/E_{\rm{CRSF}}<0.045$, which is unusually low in comparison to
the eCRSF features seen in other luminous pulsars (see above), but is more in line with
the pCRSF features claimed in some magnetar spectra (\citealt{Ibrahim02, Tiengo13}).
The similarity of $\sigma_{\rm{CRSF}}/E_{\rm{CRSF}}$ between \ngc\ and the other
high-luminosity pulsars thus further strengthens the pCRSF interpretation for M51
ULX-8 (although the inclination effects discussed by \citealt{Meszaros85a} mean it is
still difficult to completely exclude an eCRSF).

Magnetar-level B-fields are one of the possibilities invoked to explain the extreme
luminosities seen from the ULX pulsars known prior to \ngc\ (\eg\ \citealt{Mushtukov15},
and references therein). Fields this strong suppress the electron scattering cross-section
(\citealt{Herold79}) which locally reduces the radiation pressure and raises the effective
Eddington limit. In \ngc, which has a peak luminosity of
$L_{\rm{X,peak}}\sim3\times10^{39}$\,\ergps, the inferred B-field ($B\sim10^{12}$\,G) is
too weak for significant suppression of the cross-section to occur. However, M51 ULX-8
has both a much stronger field ($B\sim7\times10^{14}$\,G) and a higher peak luminosity
of $L_{\rm{X,peak}}\sim10^{40}$\,\ergps. This is similar to the other known ULX pulsars,
which also have $L_{\rm{X,peak}}\geq10^{40}$\,\ergps, $\sim$100$\times$ (or more)
the classical Eddington limit for a standard neutron star (\ie assuming the Thomson
cross-section). It is important to note that the cross-section can only be suppressed for
electrons in the regions of strong magnetic field, \ie\ the accretion column.
Phase-resolved analysis of the ULX pulsars \p13\ and \ulx1\ implies there are additional,
non-pulsed components that likely arise from the accretion flow beyond the
magnetospheric radius (\rmag) and can make a significant contribution to the total X-ray
flux (up to $\sim$50\%; \citealt{Walton18p13, Walton18ulxBB}). These cannot be subject
to the same magnetic effects, which means there is a limit to how much the B-field can
help to increase the total luminosity in these cases, and that super-Eddington accretion
is still required in addition to any magnetic effects (see also \citealt{King17ulx}).
Nevertheless, taken at face value, the results for M51 ULX-8 and \ngc\ are consistent
with the idea that in the more luminous ULXs very strong B-fields help to boost the
observed luminosity to some degree, but this does not occur in the less luminous
systems.

Assuming that \m82, \p13\ and \ulx1\ do host magnetar-level fields similar to M51 ULX-8,
given their spin periods of $\sim$1\,s it has been suggested that these fields may have to
be quadrupolar in nature in order to prevent these sources being persistently in the
propeller regime (which is clearly not the case; \citealt{Israel17}). If the difference in
B-field strength between \ngc\ and the other ULX pulsars is driven by the absence of a
strong quadrupole B-field component in the former, this may also help to explain the
much higher pulse fractions seen in \ngc\ even at the highest energies probed by \nustar\
(where dilution from the accretion flow beyond  \rmag\ is negligible; $>$70\% in \ngc\ vs
$\sim$30\% in the other ULX pulsars), as quadrupolar field geometries can result in
significantly diluted pulse fractions (\citealt{Long08}). Significantly stronger fields in \m82,
\p13\ and \ulx1\ would also naturally explain the lack of evidence for similar CRSFs in
their pulsed spectra, as they would be outside of the band currently observable.

Observationally confirming that the complexity in the pulsed spectrum of \ngc\ arises
from a CRSF is clearly of significant importance. Distinguishing between the
high-energy powerlaw tail and CRSF solutions may require extending the coverage of
the high-energy continuum to energies above 40\,keV, where the two models naturally
diverge, or may be possible by identifying the expected luminosity-dependent
variations in $E_{\rm{CRSF}}$ (\eg\ \citealt{Fuerst14vela} and references therein).
Unfortunately, based on the current peak flux exhibited by \ngc, the former will be
extremely challenging for current facilities (\ie \nustar), and may require the next
generation of hard X-ray observatory such as the \textit{High Energy X-ray Probe}
(\textit{HEX-P})\footnote{\href{https://pcos.gsfc.nasa.gov/physpag/probe/HEXP_2016.pdf}
{https://pcos.gsfc.nasa.gov/physpag/probe/HEXP\_2016.pdf}}, a potential successor to
\nustar.

%\pagebreak
%\vspace*{0.1cm}
\section*{ACKNOWLEDGEMENTS}

The authors would like to thank the reviewer for their timely and positive feedback.
DJW and MJM acknowledge support from STFC Ernest Rutherford fellowships.
ACF acknowledges ERC Advanced Grant 340442, PK acknowledges support from
the STFC, and DB acknowledges support from the French Space Agency (CNES).
This research has made use of data obtained with \nustar, a project led by Caltech,
funded by NASA and managed by NASA/JPL, and has utilized the \nustardas\
software package, jointly developed by the ASDC (Italy) and Caltech (USA). This
work has also made use of data obtained with \xmm, an ESA science mission with
instruments and  contributions directly funded by ESA Member States. 

{\it Facilites:} \facility{NuSTAR}, \facility{XMM}

\bibliographystyle{/Users/dwalton/papers/mnras}

\bibliography{/Users/dwalton/papers/references}

\label{lastpage}

\end{document}